# Hiding Information in Big Data based on Deep Learning

Dingju Zhu[1,2] *

1. School of Computer Science, South China Normal University, Guangzhou, China
2. School of Artificial Intelligence and Robot Education Industry, South China Normal University, Guangzhou, China
Corresponding Author's Email: zhudingju@m.scnu.edu.cn

**Abstract:** The current approach of information hiding based on deep learning model can not directly use the original data as carriers, which means the approach can not make use of the existing data in big data to hiding information. We proposed a novel method of information hiding in big data based on deep learning. Our method uses the existing data in big data as carriers and uses deep learning models to hide and extract secret messages in big data. The data amount of big data is unlimited and thus the data amount of secret messages hided in big data can also be unlimited. Before opponents want to extract secret messages from carriers, they need to find the carriers, however finding out the carriers from big data is just like finding out a box from the sea. Deep learning models are well known as deep black boxes in which the process from the input to the output is very complex, and thus the deep learning model for information hiding is almost impossible for opponents to reconstruct. The results also show that our method can hide secret messages safely, conveniently, quickly and with no limitation on the data amount.

## 1. Introduction

Deep learning has been used in many applications including disease treatment[1], resource management[2] and so on. Big data and cloud data bring many new challenges to many fields[3], especially to information security[4,5].

Information hiding is a way to hide confidential information in a large amount of information and not to let the opponent find it. The data amount of the carrier is limited in the existing technology of information hiding, which means the data amount of secret message is limited for it is impossible to hide more data amount of secret message than the data amount of the carrier. If we make use of the data in big data as carriers, we can not only solve this problem, but also can make information hiding more security, for before opponents want to extract secret messages from carriers, they need to find the carriers, however finding out the carriers from big data is just like finding out a box from the sea.

The algorithm for information hiding is as important as the carrier. Deep learning is equivalent to a complex algorithm that cannot be explained explicitly. Deep learning models are well known as deep black boxes in which the process from the input to the output is very complex. If the deep learning model is used for information hiding, it may achieve unexpected results. Deep learning has been used to detect image steganography in some researches [6-10], in which the deep learning model is used to determine whether or not an image is a steganographic image and can not be used to hide information. Deep learning has also been used to hide information in few researches.

In the first existing method of hiding information based on deep learning, the deep learning model is used to encode the cover carrier and secret message into a stego carrier in the process of hiding information and decode the secret message from the stego carrier in the process of extracting information. If the cover carrier is the data stored in big data, the data has to be modified to the stego carrier, which will destroy the data in big data, or the stego carrier has to be insert into big data as new data, which can not make use of the existing data in big data and will waste storage resources. In the second existing method of hiding information based on deep learning, the deep learning model generates a cover carrier into which the secret message will be inserted to form the stego carrier, and thus the generated carriers are not the data in big data and will waste storage resources. In the third existing method of hiding information based on deep learning, the deep learning model uses the secret message to generage the stego carrier from which the secret message can be extracted[11]. The stego carrier is not the existing data in big data and thus has to be inserted into big data, which will waste storage resources. There the existing methods are not suitable for hiding information in big data for not using the data in big data as the carrier.

In our method, the original carrier and secret message are used as the input and expected output to train the deep learning model, the original carrier is inputted into the trained deep learning model to obtain the output result which is usually close to and has a little difference with the secret message, and then the original carrier and little difference can be used to recover the secret message by inputting the original carrier into the deep learning model to obtain the output result which can be used to recover the secret message. Our method can make use of the existing data in big data as the carrier and recover the secret message from the carrier based on deep learning, and thus does not need to modify the data in big data or insert new data into big data.

There are some researches on information security in big data[12], however there is no existing research on hiding information in big data except a related research about hiding the secret message in a webpage that contains all the information to describe the secret message. However, the related research can only be used to hide text messages, cannot be used to hide other types of messages such as images, audios and videos, and when the secret message is larger than a web page, it is impossible to find a webpage that contains all the information to describe the secret message. What is more, the hiding algorithm of the related research is just a mapping between the text of the secret message and the text of the web page, so the secret message is easily extracted from the web page. Our method is based on big data which can hide various types of secret messages with no limit on the data amount, and based on deep learning which can make the secret message very difficult to extract.

Compared with the existing methods, our method of information hiding in big data based on deep learning has three advantages. Firstly, the data in big data will not be destroyed; secondly, no more storage space is occupied; thirdly, the carrier is difficult to find by the opponent for the carrier is the existing data in big data, and if the carrier can not be found, the secret information is certainly impossible to be found; fourthly, the secret message with an infinite data amount can be hided for the carrier are from big data; finally, the deep learning model is very complicated and difficult to explain, so the hiding algorithm based on deep learning is hard to break by the opponent.

The main contributions of this paper includes:

(1) The data in big data is used as the carrier for information hiding, which makes the carrier almost impossible to find by the opponent, and also saves space for storing the carrier.

(2) The training and testing process of deep learning model are used as the algorithm of hiding and extracting process, which makes the secret message almost impossible to be recovered for the extreme complexity of deep learning model.

(3) Our method is unbreakable when the sender and receiver share the deep learning model not through network.

The rest of the paper is organized as follows. Section 2 reviews related work. Section 3 describes methodology. Section 4 shows experiments and comparisons. Then Section 5 presents the results and discussion. Finally, we conclude this paper in Section 6.

## 2 Review of related work

There are a lot of paper provide methods for detecting image steganography based on deep learning[6-10,13-26]. The deep learning models are used to determine whether the image to be detected is a steganographic image. However the methods can only be used for detecting image steganography, and can not be used for information hiding. Our paper provides a method for information hiding based on deep learning model.

Generating steganographic images via adversarial training [27], hiding images in plain sight [28] and HiDDeN[29] researched on hide data with deep networks；deep neural networks for speech steganography [30] proposed the use of neural networks for speech steganography. The main idea of them is to get as input the carrier and the message to generate the new encoded carrier, and then decode the message from the new encoded carrier. If the carrier is in big data, then the approach need to revise or copy the carrier in the big data to store the new encoded carrier. However, the big data is generally not allowed to be modified, and copying will make the big data increase dramatically. And thus the method is not suitable when the carrier is in big data. The new idea of our paper is to get as input the carrier in big data and expected output the message to train the model, and then recover the message from the carrier directly by the model, which does not need to modify the carrier, and thus is very suitable for big data.

Secure steganography based on generative adversarial networks (SSGAN) proposed the generation of images cover and then use them to make insertion by modification[31]. Chaumont said the method seems to complicate the matters for it is more logic to choose natural images rather than generating images and then using them to hide a message[32]. Another image steganography method via deep convolutional generative adversarial networks produces the cover images according to the secret information by generator without modification and use extractor to extract information from stego images produced by generator[11]. The methods are also not suitable for big data for the generated images will occupy the space in big data, and whatsmore, the methods can not make use of the existing data such as images in big data. Our method can make use of the existing data such as images in big data and thus no more space is occupied, and our method only need one model rather than two models including the generator and the extractor.

Embedding watermarks into deep neural networks protect the rights to shared trained models[33]. The method regards shared trained models as covers and has no relations with make use of the deep learning model to detect or hide secret information.

It can be seen that the existing methods of hiding using neural networks either need to modify the carrier or create a new carrier. Modifying the carrier is not feasible for big data, because third-party big data often does not allow us to modify it. Creating a new carrier will generate new data which take up new storage space and cost more than using existing carriers. Our method can make use of the existing data in big data as the carrier and hide the information through the neural network including the deep learning model.

There are a lot of researches about information security in big data[12], however only one research related with big data for hiding information or steganography is found. The approach to text steganography based on search in internet[34] can hide the secret message in a webpage that contains all the information to describe a secret message. However, the method can only be used to hide text information, and cannot be used for hiding various other types of information such as images, audio, and video. And the method assumes that the secret message is relatively small, so when the secret message is complicated or large, for example when the secret message is larger than a web page, it is difficult to find a webpage that contains all the information to describe the secret message, then the method will fail. Our method is applicable to the hiding of various types of information, and there is no limit to the size of the secret message in our method.

## 3. Methodology

### 3.1 Basic method of information hiding in big data based on deep learning

As shown in Figure 1, the method of hiding information is used in conjunction with the method of extracting information.

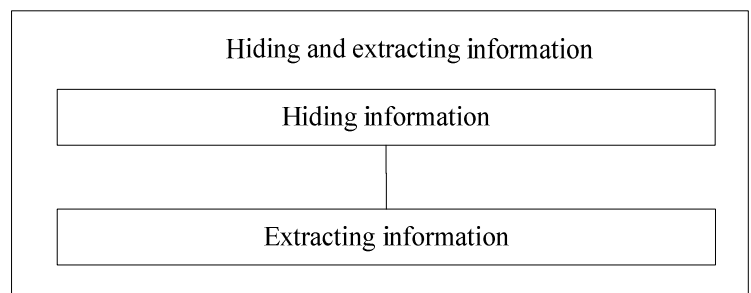

**Figure. 1** Hiding and extracting information .

**(1)Hiding information in big data based on deep learning**

As shown in Figure 2, the process of hiding information in big data based on deep learning includes accepting the secret message, selecting the cover carrier, training the deep learning model, generating the stego carrier, generating the secret difference, hiding information multiple times, generating the authorization and verification information, and transmitting the authorization and verification information.

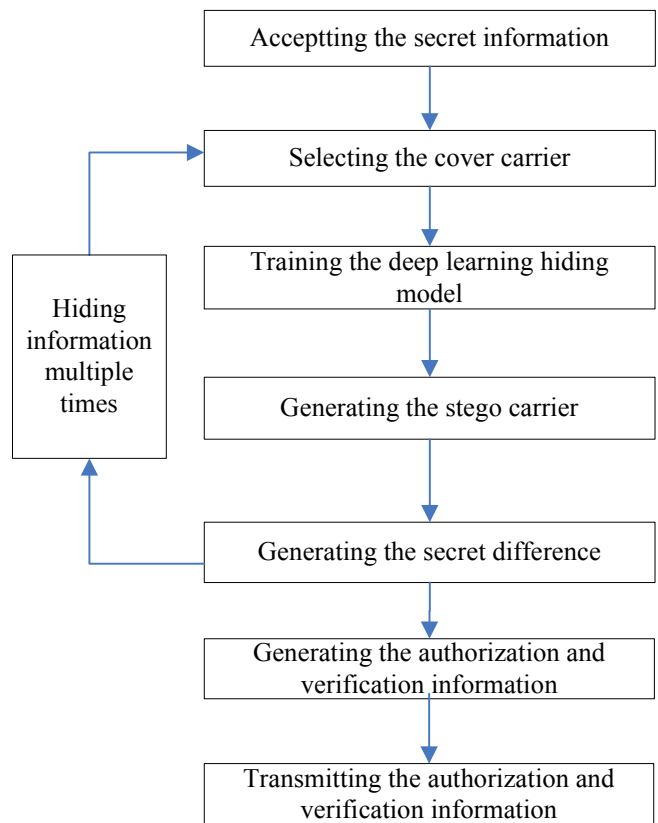

**Figure.2** Hiding information .

The secret message inputted by the sender is accepted. The secret message refers to the information that the sender needs to hide, which can be uploaded by the sender through a browser, a client or the like. For example, the secret message contains 100 sentences. The format of the secret message can be in various formats such as text, audio and video.

The data in big data is selected as a cover carrier, and the location of the data in big data is obtained as the secret location. The carrier is selected according to the secret message or a preset rule such as selecting a consecutive block whose size is K from an inactive file in big data. If the cover carrier is consecutive, the secret location includes the start and end positions of the cover carrier; if the cover carrier is inconsecutive, the secret location includes multiple sets of start and end positions of consecutive data segments.

The data with the same type of the cover carrier and the message with the same type of the secret message are used as the input and expected output to train the deep learning model for information hiding, and the trained deep learning model is called the deep learning hiding model. The deep learning model can be CNN ,RNN and so on. The deep learning model is a kind of artificial neural network with little difference between the expected output and the actual output after training. If the computing ability is sufficient, the deep learning model is preferred, but if the computing ability is not enough, the traditional artificial neural network such as BP neural network can also be used instead of the deep learning model in our method. If the trained deep learning hiding model has existed, the model can be used directly.

The cover carrier in big data is inputted into the deep learning hiding model, and the output of the deep learning hiding model is used as the stego carrier. Since the deep learning hiding model has been trained by the data with the same type of the secret message as the expected output, when the cover carrier is input, the output of the deep learning hiding model has certainly the same type of the secret message.

The difference between the secret message and the stego carrier is computed and taken as the secret difference. The stego carrier has the same type of the secret message, and thus the stego carrier can be compared with the secret message. The formula for computing the secret difference is the secret difference = f (the secret message, the stego carrier), where f is a function for computing the difference, and if the secret message and stego carrier are both numbers, the formula can be simplified as the secret difference = secret message - the stego carrier.

The above processes can be executed multiple times. Multiple cover carriers and secret locations are obtained, and multiple deep learning hiding models are trained by using the input data with the same type of the cover carriers and the expected output data with the same type of the secret message. The differences between the secret message and the stego carriers are computed to obtain multiple secret differences.

The authorization information includes at least one set of secret difference, secret location, and parameters of the deep learning hiding model. The parameters of the deep learning hiding model include the type of neural network, number of layers, number of nodes per layer, weight of each connection between nodes in each layer, and so on. If the secret location is sent to the receiver, the receiver can obtain the cover carrier from the secret location in big data. If the deep learning hiding model is sent to the receiver, the receiver can input the cover carrier into the deep learning hiding model to compute out the stego carrier. If the secret difference is sent to the receiver, the user can synthesize the secret message according to the stego carrier and the secret difference. The secret difference, secret location, and deep learning hiding model are indispensable, and the receiver is impossible to extract the secret message without any one of them.

The partial data or the attribute information of the secret message is taken as the verification information and sent to the receiver. For example, the data in the last m bytes of the secret message or the number of bytes of the secret message is used as the verification information. The receiver can verify whether or not the extracted secret message is consistent with the verification information, which is very necessary for the cover carrier in big data may change which will lead to the extracted secret message is wrong. When there are multiple sets of authorization information, the secret message consistent with the verification information can be selected from the multiple extracted secret messages.

The authorization and verification information are sent to the receiver.

**(2) Extracting information in big data based on deep learning**

As shown in Figure 3, the process of extracting information in big data based on deep learning includes accepting the authorization information, extracting the cover carrier, generating the stego carrier, generating the secret message, extracting information multiple times, synthesizing the secret message, and transmitting the secret message.

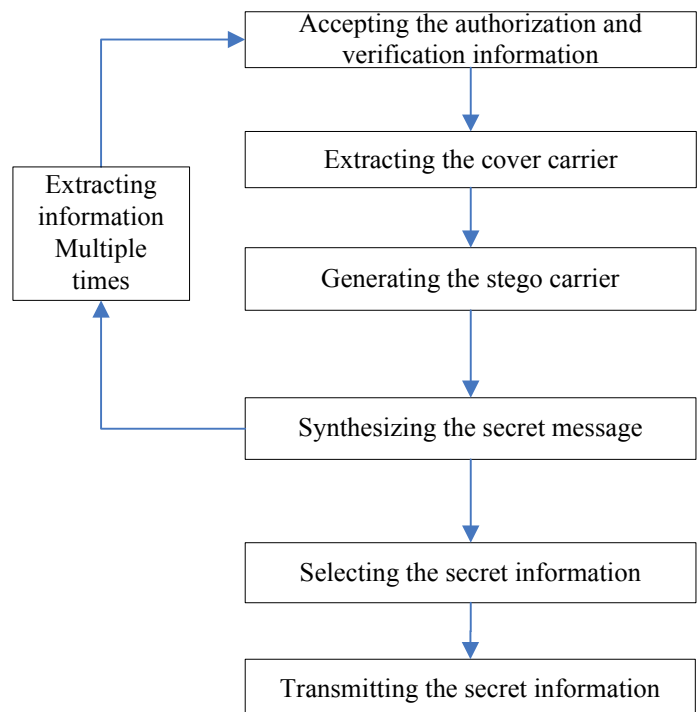

**Figure. 3** Extracting information.

The verification information and at least one set of the authorization information including the secret difference, secret location, and parameters of the deep learning hiding model inputted by the receiver are accepted. The deep learning hiding model is automatically generated and configured according to the parameters.

The cover carrier is obtained from big data according to the secret location. The cover carrier is inputted into the deep learning hiding model, and the output is used as the stego carrier. The secret message is synthesized according to the stego carrier and the secret difference. When recovering the secret message, the secret difference and the stego carrier are substituted into the formula the secret difference = F (secret message, the stego carrier) to calculate the secret message. If the secret difference and the stego carrier are both numbers, the formula can be simplified as the secret message = the stego carrier + the secret difference.

The multiple sets of authorization information may be used to obtain multiple secret messages by executing the above process multiple times. The secret message that is consistent with the verification information is selected from the multiple secret messages and transmitted to the receiver. The receiver can also view or download online through a browser, a client or the like.

### 3.2 Parallel method of information hiding in big data based on deep learning
Parallel hiding and extracting is very necessary for the secret message which contains a very large amount of data, which can greatly speed up the process of hiding and extracting information

**(1) Hiding information in big data based on deep learning in parallel**
As shown in Figure 4, the process of hiding information in parallel is based on the process of hiding shown in Figure 2. Accepting the secret message further includes dividing the secret message; selecting the cover carrier further includes selecting multiple cover carriers; training the deep learing hiding model further includes training deep learing hiding models; generating the stego carrier further includes generating multiple stego carrieres; generating the secret difference further includes generating multiple secret differences; generating the authorization and verification information further includes generating multiple sets of authorization and verification information, and transmitting the authorization and verification information further includes transmitting multiple sets of authorization and verification information

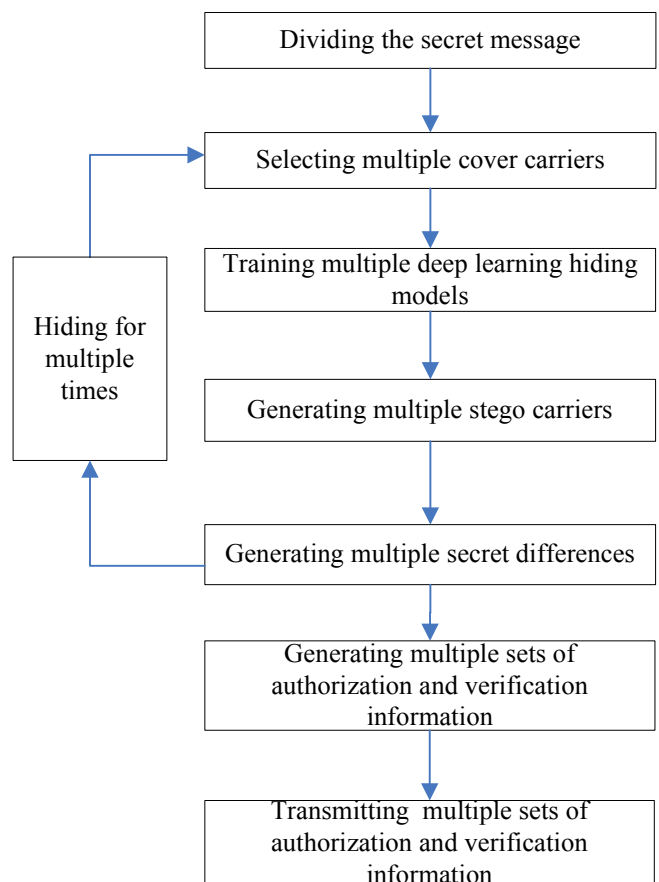

**Figure. 4** Hiding information in parallel.

The secret message is split into P secret sub-messages, where P>=1, and the position number of each secret sub-messages in the secret message is called as the secret number. For example, there are 100 sentences in the secret message and each sentence is divided into 100 strings. The position number of a string is 004006, which means the string is the sixth string in the fourth sentence. Hiding secret message after segmentation can furtherly improve the security of information hiding, for the opponent can not recover the whole secret message even if he recovers parts of the secret message.

P cover carriers are selected from big data, and one-to-one correspondences between the P cover carriers and the P secret sub-messages are erected. The P cover carriers are located at P locations which can be storage locations or paths or index numbers. The locations are obtained as the secret locations.

The data with the same type of each of the cover carriers and the data with the same type of its corresponding secret submessage are used as the input and expected output to train its corresponding deep learning hiding model, and multiple deep learning hiding model are obtained. If the trained deep learning hiding models has existed, then the model can be used directly. Each of the cover carriers is inputted into its corresponding deep learning hiding model, and the output is obtained as its corresponding stego carrier.

There are one-to-one correspondences between the secret sub-messages and the stego carriers, for there are one-to-one correspondences between the cover carrieres and the secret sub-messages, and there are one-to-one correspondences between the cover carriers and the stego carriers. The difference between each of the secret sub-messages and its corresponding stego carrier is taken as its corresponding secret difference.

The multiple sets of secret location, secret numbers, secret difference, and deep learning hiding model are transmitted to the receiver, which can be used to generate the multiple secret sub-messages during the process of extracting information.

**(2) Extracting information in big data based on deep learning in parallel**

As shown in Figure 5, the process of extracting information in parallel is based on the process of extracting shown in Figure 3. Accepting the authorization and verification information further includes accepting multiple sets of authorization and verification information; extracting the cover carrier further includes extracting multiple cover carriers; and generating the stego carrier further includes generating multiple stego carriers; synthesizing the secret message further includes synthesizing multiple secret sub-messages and synthesizing the secret message.

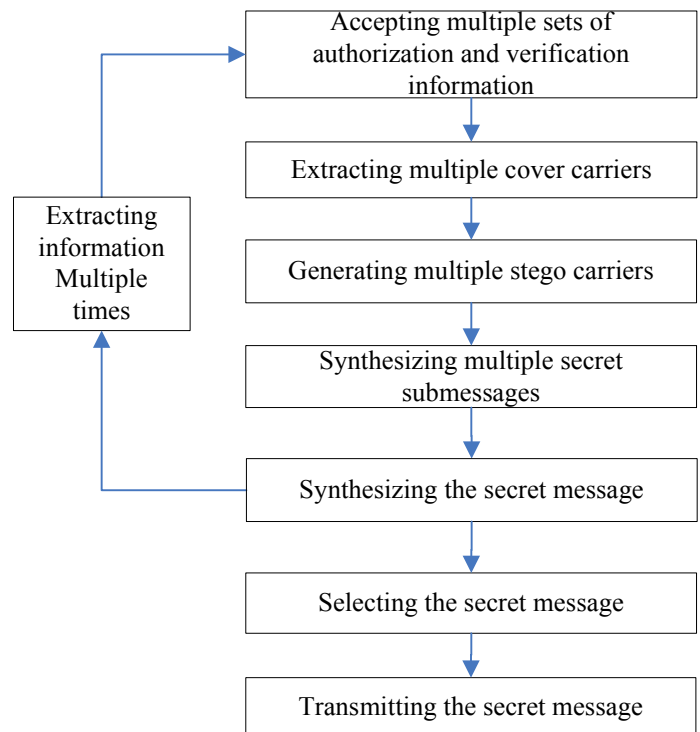

**Figure. 5** Extracting information in parallel.

Multiple sets of secret location, secret numbers, secret difference, and deep learning hiding model inputted by the receiver are accepted. Each of the cover carrier from big data is located and obtained through the secret locations. Each of the cover carriers is inputted into its corresponding deep learning hiding model, and the output is taken as its corresponding stego carrier. Each of the secret sub-messages is synthesized according to its corresponding stego carrier and secret difference. The secret message is synthesized according to the secret sub-messages and the secret numbers.

**3.3 Unbreakable method of information hiding in big data based on deep learning**

The only approach to break our method is that the opponent obtains the authorization information and knows how to use our method to recover the secret message based on the authorization information. If not all parts of the authorization information are obtained by the opponent, the opponent is impossible to recover the secret message for any part of the authorization information is indispensable for recovering the secret message. Different parts of the authorization information can be sent to the receiver through different channels at different times, which can greatly increases the difficulty of all parts of the authorization information being obtained by the opponent.

In the process of hiding information, the secret difference can not be generated without the secret message, the secret location is not dependent on the secret message, and the deep learning hiding model is not dependent on the content of the secret message but the type of the secret message. If the type of the secret message is known ahead of time, the deep learning hiding model can be generated ahead of time, and the deep learning hiding model can be used for hiding all kinds of secret messages with the type. Therefore, the sender and receiver can share the secret location or the deep learning hiding model ahead of time not through network but other approaches such as copy or express delivery. Then the sender and receiver can use the secret location or the deep learning hiding model to hide and extract the secret message. Since only the secret difference are transmitted on the network between the sender and the receiver, even if the opponent intercepts the secret difference, the secret message cannot be recovered at all without the secret location or the deep learning hiding model, and thus the method is unbreakable for information hiding.

Even if the secret location or the deep learning hiding model have to be shared between the sender and the receiver through network, there are two methods to reduce the risk of network transmission being intercepted. The first method is that the secret location or the deep learning hiding model is transmitted long before the secret difference is transmitted. If the interval is enough long, the opponent will think the secret location or the deep learning hiding model has no relation with the secret difference. The second method is that the transmission of different parts of the authorization information in different directions. For example, the receiver sends the deep learning hiding model and secret location to the sender, and then the sender send the secret difference to the receiver. The two methods can extremely reduce the possibility of the opponent to intercept and use together the different parts of the authorization information.

## 4. Experiments and comparisons
### 4.1 Experiments

The computer we used for hiding and extracting information has a 3.80GHZ CPU and a 4G memory, and the trained deep learning hiding model we used can accept an input image and output a text. The process of hiding information with the trained deep learning hiding model takes only about 3 seconds, and the process of extracting information also takes only about 3 seconds.

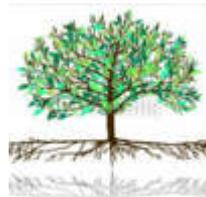

**Figure. 6** Image in big data from the Internet.

Experiment 1: The sender Alice needs to send "Knife" as the secret message to the receiver Bob.

The process of hiding information: The secret message "Knife" is inputted by Alice, and an image in big data from the Internet is randomly downloaded. The image is shown in Figure 6, and the location of the image in big data is the url https://origin-thumbs.bigstockphoto.com/zh/image-4671121/stock-vector-树. A trained deep learning model which can recognize more than one thousand objects is obtained as the deep learning hiding model. The image is inputted into the deep learning hiding model, and the output is "pot, flowerpot " with a score of 0.74342. From the output we can see that the recognition result is actually wrong, but it does not affect the effect of the experiment. By converting Ascii text to decimal, the output "pot, flowerpot" is converted to the array [112, 111, 116, 44, 32, 102, 108, 111, 119, 101, 114, 112, 111, 116] and the secret message is converted into the array [107, 110, 105, 102, 101]. The lengthes of the two arrays are different, and thus in order to comparing the difference, the array [112, 111, 116, 44, 32, 102, 108, 111, 119, 101, 114, 112, 111, 116] is converted to the array [112, 111, 116, 44, 32] with the same length of the array of the secret message. The difference between the two arrays is calculated to obtain the array of the secret difference [-5, -1, -11, 58, 69]. The location of the image in big data, the parameters of the deep learning hiding model, and the array of the secret difference are sent to Bob through different channels.

The process of extracting information: Bob receives the location of the image in big data, the parameters of the deep learning hiding model and the array of the secret difference. The deep learning hiding model is reconstructed according to the parameters. The image is downloaded from the location and inputted into the deep learning hiding model to compute the output. The output is "pot, flowerpot " with a score of 0.74342. The output "pot, flowerpot " is converted into the array [112, 111, 116, 44, 32, 102, 108, 111, 119, 101, 114, 112, 111, 116] which needs to be converted to the array [112, 111, 116, 44, 32] with the same length of the array of the secret difference. The array of the output is added to the array of secret difference to get the array [107, 110, 105, 102, 101] which is converted into the secret message "Knife" by decoding decimal number to ascii text .

Experiment 2: The sender Alice needs to send the secret message "We will meet at the place we met last week at 12 o'clock tomorrow morning." to the receiver Bob.

The process of hiding information: the secret message "We will meet at the place we met last week at 12 o'clock tomorrow morning " is inputted by Alice. The image and the deep learning hiding model is same as the first experiment. The output of the deep learning hiding model is converted to the array [112, 111, 116, 44, 32, 102, 108, 111, 119, 101, 114, 112, 111, 116], and the secret message is converted into the array [87, 101, 32, 119, 105, 108, 108, 32, 109, 101, 101, 116, 32, 97, 116, 32, 116, 104, 101, 32, 112, 108, 97, 99, 101, 32, 119, 101, 32, 109, 101, 116, 32, 108, 97, 115, 116, 32, 119, 101, 101, 107, 32, 97, 116, 32, 49, 50, 32, 111, 39, 99, 108, 111, 99, 107, 32, 116, 111, 109, 111, 114, 114, 111, 119, 32, 109, 111, 114, 110, 105, 110, 103, 46]. In order to comparing the difference, the array [112, 111, 116, 44, 32, 102, 108, 111, 119, 101, 114, 112, 111, 116] is converted to the array [112, 111, 116, 44, 32, 102, 108, 111, 119, 101, 114, 112, 111, 116, 112, 111, 116, 44, 32, 102, 108, 111, 119, 101, 114, 112, 111, 116, 112, 111, 116, 44, 32, 102, 108, 111, 119, 101, 114, 112, 111, 116, 112, 111, 116, 44, 32, 102, 108, 111, 119, 101, 114, 112, 111, 116, 112, 111, 116, 44, 32, 102, 108, 111, 119, 101, 114, 112, 111, 116, 112, 111, 116, 44] with the same length of the array of the secret message. The difference between the elements of two arrays is calculated to get the difference array [-25, -10, -84, 75, 73, 6, 0, -79, -10, 0, -13, 4, -79, -19, 4, -79, 0, 60, 69, -70, 4, -3, -22, -2, -13, -80, 8, -15, -80, -2, -15, 72, 0, 6, -11, 4, -3, -69, 5, -11, -10, -9, -80, -14, 0, -12, 17, -52, -76, 0, -80, -2, -6, -1, -12, -9, -80, 5, -5, 65, 79, 12, 6, 0, 0, -69, -5, -1, 3, -6, -7, -1, -13, 2]. The location of the image in big data, the parameters of the deep learning hiding model, and the array of the secret difference are sent to Bob through different channels.

The process of extracting information: Bob receives the location of the image in big data, the parameters of the deep learning hiding model, and the array of secret difference. The image is downloaded from the location in big data, and inputted into the deep learning hiding model to compute the output. The output is "pot, flowerpot " with a score of 0.74342. The output is converted into the array [112, 111, 116, 44, 32, 102, 108, 111, 119, 101, 114, 112, 111, 116] which needs to be converted to the array [112, 111, 116, 44, 32, 102, 108, 111, 119, 101, 114, 112, 111, 116, 112, 111, 116, 44, 32, 102, 108, 111, 119, 101, 114, 112, 111, 116, 112, 111, 116, 44, 32, 102, 108, 111, 119, 101, 114, 112, 111, 116, 112, 111, 116, 44, 32, 102, 108, 111, 119, 101, 114, 112, 111, 116, 112, 111, 116, 44, 32, 102, 108, 111, 119, 101, 114, 112, 111, 116, 112, 111, 116, 44] with the same length of the array of the secret difference. The array of the output is added to the array of secret difference to get the array [87, 101, 32, 119, 105, 108, 108, 32, 109, 101, 101, 116, 32, 97, 116, 32, 116, 104, 101, 32, 112, 108, 97, 99, 101, 32, 119, 101, 32, 109, 101, 116, 32, 108, 97, 115, 116, 32, 119, 101, 101, 107, 32, 97, 116, 32, 49, 50, 32, 111, 39, 99, 108, 111, 99, 107, 32, 116, 111, 109, 111, 114, 114, 111, 119, 32, 109, 111, 114, 110, 105, 110, 103, 46] which is converted into the secret message " We will meet at the place we met last week at 12 o'clock tomorrow morning." by decoding decimal number to ascii text .

### 4.2 Comparison with existing work

The existing methods for detecting image steganography based on deep learning[6-10,13-26] can not be used in hiding secret messages such as "Knife" or "We will meet at the place we met last week at 12 o'clock tomorrow morning." shown in our experiments. And the deep learning models in the existing methods in [6-10,13-26] can be used to determine whether the image to be detected is a steganographic image in most cases, but can not work for the image used in our method. For example, although the secret messages are generated from the tree image by our deep learning model, the tree image used in our experiments is actually without any secret messages, and thus the deep learning models in the existing methods in [6-10,13-26] can not detect image steganography. This illustrates from another aspect that the information hidden by our method is difficult to detect by the existing methods.

The existing methods such as generating steganographic images via adversarial training [27], hiding images in plain sight [28], HiDDeN[29] and deep neural networks for speech steganography [30] can hide data with deep networks or neural networks by getting as input the carrier and the message to generate the new encoded carrier, and then decoding the message from the new encoded carrier. Comparing with our experiments, the existing methods need to get the tree image and the secret messages such as "Knife" or "We will meet at the place we met last week at 12 o'clock tomorrow morning." together as the input of the deep learning model for encoding, and the output of the deep learning model will be a new tree image hiding the secret messages "Knife" or "We will meet at the place we met last week at 12 o'clock tomorrow morning." And then, the secret messages "Knife" or "We will meet at the place we met last week at 12 o'clock tomorrow morning." can be decoded by the deep learning model for decoding. Although the existing methods in [27-30] can hiding secret information, the new tree image hiding the secret messages can be detected as image steganography by the existing methods in [6-10,13-26], however the tree image used for hiding the secret messages in our method can not be detected by the existing methods in [6-10,13-26], and thus the secret messages hided by our method are more difficult to be detected than the existing methods in [27-30]. What' more, in our experiment, the tree image is obtained from the internet big data, if we use the existing methods in [27-30], we have to revise the tree image in the big data to the new tree image containing the secret messages or copy the new tree image into the big data to store the new encoded carrier. However, the internet big data is generally not allowed to be modified, for the tree image is obtained through a URL, and we have no permission to modify the content of the image corresponding to the URL. And copying the new tree image into the Internet big data and generate a new URL also need the permission and it is impossible for ordinary users to have such permission. Even if we have the permission, the copied new tree image will make the big data increase dramatically when there are many new carriers are copied into the big data, which is not good for saving big data storage resources. Our methods as shown in our experiments get the tree image obtained from big data as input of our deep learning model to hide the secret messages and then can decode the secret messages by use the same tree image as the input of our deep learning model, and thus no more new carriers are generated. So our method does not require any modification to the carrier in big data or write any carrier into big data. Therefore, our method can effectively use big data resources without causing any damage and pressure on big data resources. However the existing methods in [27-30] can not work when the big data is not permitted to change.

Secure steganography based on generative adversarial networks proposed the generation of image coverers for insertion of secret information[31]. Another image steganography method via deep convolutional generative adversarial networks produces the cover images according to the secret information [11]. Although the existing methods in [31] and [11] have not used of the tree image in the big data as shown in our experiments, the existing methods in [31] and [11] have also generated the new image as the new coverer which has to be copied into big data, which will lead to the waste of big data storage space, and thus the existing methods in [31] and [11] is essentially the same as the existing methods in [27-30]. Moreover, the methods in [31] and [11] are not as good as those existing methods in [27-30] for they cannot make use of existing images in big data, but generate new images. The concealment of such generated new images are not as good as using existing images in big data, because existing images in big data are organic parts of big data and hard to be detected.

If the generated images in these existing methods in [11, 27-31] are not put into big data, although they will not cause changes or growth of big data, they will certainly occupy the user's own storage space. When the amount and size of the generated images are very large, this problem becomes very prominent, but these existing methods in [11, 27-31] can not rely on and use the data in the existing big data, so they will be powerless on this problem. These shortcomings are exactly the advantages of our method. At the same time, if the images are stored directly on the client side, then the images need to be transmitted between the clients, which will put pressure on the network between the clients, and it is easier to attract attackers' attention to detect the hidden information from the images. Our method only needs to transfer the addresses of the images in big data, the addresses are only a very small amount of data, and it is not easy to be noticed by attackers. Even if the addresses are intercepted and the images are obtained from the big data, no secret information can be obtained from the images because the images themselves do not contain any secret information as shown in our experiments.

Embedding watermarks into deep neural networks protect the rights to shared trained models[33]. The existing method regards shared trained models as covers and actually has no relations with our method which makes use of the deep learning model to detect or hide secret information.

There are a lot of existing researches about information security in big data[12], however only one research slightly related with big data for hiding information or steganography is found. The approach to text steganography based on search in Internet [34] hiding the secret message in a webpage that contains all the information to describe a secret message . The existing method in [34] can only be used to hide text information, and cannot be used for hiding various other types of information such as images, audio, and video, because it is difficult to find images, videos, audios, etc. that are the same as the parts of secret information from the web pages. Even considering only hidden text, the existing method in [34] can work only in limited situations. The existing method in [34] assumes that the secret message is relatively small, so it's not difficult to find a web page that contains all the parts of the information. However, if the secret message is complicated or large, or only a few web pages can be obtained, which means that the access rights of the web pages are limited, then the existing method in [34] may not be able to find the components of the secret text from the web pages. What's more, this existing method does not process the information to be hidden through a deep learning model, so it is easier to detect then our method. None of these shortcomings of this existing method is a problem for our method. Although our experiments are just examples of text secret information, our method is also applicable to data in other formats, such as images, video, audio, and so on, because our method uses the output of the deep learning model as secret information, and the output format of a deep learning model can be either text, or other formats such as images or video or audio. For example, deep learning models such as unet and GAN can generate images or video or audio.

## 5. Results and discussions

Although the secret message in our experiments is text, our method can also be used to hide the secret message with the types of image, video, audio,and so on by changing the output type of the deep learning hiding model.

In the two experiments, even if the opponent obtains the image in big data, when he see a tree in the image, so it is almost impossible for him to think that the secret message may be "Knife" or "We will meet at the place we met last week at 12 O'clock tomorrow morning"; even if the opponent obtains the parameters of the deep learning hiding model, from which he knows the object types which the model can recognize, he does not know which type is related with the secret message and almost impossible to guess the type for there are more than one thousand types in the parameters.

The secret location and the parameters of deep learning hiding model seems to have no relationship with the secret message, so even if the opponent interceptes the authorization information, the secret message could not be recovered without knowing the role of each part of the authorization information. The cover carrier is selected from big data, so it is very difficult for the opponent to find the cover carrier for big data is too large. Even if the cover carrier is found, the opponent can not get the stego carrier without the deep learning hiding model. Further, even if the stego carrier is obtained, the secret message cannot be recovered without the secret difference. The three insurmountable thresholds make our method extremely hard to break, and when the secret location or the deep learning hiding model is shared not through network before information hiding, our method is impossible to break by the opponent.

Big data is often stored in cloud with many users, which means that it is difficult to ensure the data in big data does not change. If the cover carrier changes in big data, the stego carrier outputted by the deep learning hiding model in the process of extracting information will be not consistent with the stego carrier in process of hiding information, and finally the recovered secret message will be wrong. Our method makes use of multiple cover carriers to overcome this problem. As long as one of the cover carriers does not change, the secret message can always be recovered correctly with the help of the verification information, which can greatly improve the robustness of our method.

If the data amount of the secret message is very large, the secret message can be divided into multiple secret sub-messages in our method. Since the data amount of big data is very large, the different carriers corresponding to the different secret sub-messages can be selected. Therefore, the data amount of the secret message can be unlimited.

Moreover, different channels may be used for transmitting the different parts of the authorization information, and the authorization information can be transmitted at different times. For example, the secret difference is sent to the receiver through the first network path, the secret location through the second network path, and the deep learning hiding model through the third network path, which can make the opponent harder to break our method, for the opponent can not recover the secret message even if parts of the authorization information are intercepted. Our method becomes unbreakable when the sender and the receiver share the secret location or the deep learning hiding model not through network before information hiding.

## 6. Conclusion

We proposed a novel method of information hiding in big data based on deep learning. Our method uses the existing data in big data as the carrier, and uses the deep learning model as the function of hiding and extracting information. In the process of hiding information, the cover carrier in big data is inputted into the deep learning hiding model, the output of the model is used as the stego carrier, and the difference between the stego carrier and the secret message is taken as the secret difference. In the process of extracting information, the cover carrier in big data is inputted into the deep learning hiding model, the output of the model is used as the stego carrier, and the secret message is synthesized according to the stego carrier and the secret difference. The results also show that our method can hide secret messages safely, conveniently, quickly and with no limitation on the data amount. When the secret location or the deep learning hiding model is shared not through network before information hiding, our method is unbreakable. The deep learning model in our method can also be replaced by other functions such as BP artificial neural network and big data in our method can also be replaced by other data sources such as Internet data and IOT data, which means that our method opens a new door for information hiding.

**Acknowledgements**

The authors acknowledge the National New Engineering Research and Practice Project (Grant: Higher Education Ministry Letter [2018]17); the Major Project of National Social Science Fund(14ZDB101); the Guangdong Province Higher Education Teaching Research and Reform Project (Grant: Guangdong Higher Education Letter [2016]236) ; the Guangdong Province Graduate Education Innovation Project (Grant: 2016JGXM_ZD_30 ) ; the Guangdong Province Joint Training of Graduate Demonstration Base (Grant: Guangdong Teaching and Research Letter [2016]39 ) ; the Guangdong Province New Engineering Research and Practice Project (Grant: Guangdong Higher Education Letter [2017]118 ) ; the Major Provincial Scientific Research Projects in Guangdong Universities (Grant: Guangdong Teaching and Scientific Letter [2018] 64).